\documentclass[%
reprint,
superscriptaddress,
 amsmath,amssymb,
 aps,
 pra,
]{revtex4-1}
\usepackage[dvipsnames]{xcolor}
\definecolor{mblue}{RGB}{42, 54, 144} 
\usepackage{hyperref}
\usepackage{graphicx}
\usepackage{dcolumn}
\usepackage{bm}
\usepackage{braket}
\usepackage{amssymb}
\usepackage{changes}
\usepackage{siunitx}

\begin{document}

\title{Measurement of three-body recombination coefficient of ultracold lithium and strontium atoms}

\author{Bo-Yang Wang}
\thanks{These authors contributed equally to this work.}
\affiliation{State Key Laboratory of Low-Dimensional Quantum Physics, Department of Physics, Tsinghua University, Beijing 100084, China}

\author{Yi-Fan Wang}
\thanks{These authors contributed equally to this work.}
\affiliation{State Key Laboratory of Low-Dimensional Quantum Physics, Department of Physics, Tsinghua University, Beijing 100084, China}

\author{Zi-He An}
\affiliation{State Key Laboratory of Low-Dimensional Quantum Physics, Department of Physics, Tsinghua University, Beijing 100084, China}

\author{Li-Yang Xie}
\affiliation{State Key Laboratory of Low-Dimensional Quantum Physics, Department of Physics, Tsinghua University, Beijing 100084, China}

\author{Zhu-Xiong Ye}
\email{Present address: State Key Laboratory of Quantum Optics and Quantum Optics Devices, Institute of Opto-Electronics, Shanxi University, Taiyuan, Shanxi 030006, China.}
\affiliation{State Key Laboratory of Low-Dimensional Quantum Physics, Department of Physics, Tsinghua University, Beijing 100084, China}

\author{Yi Zhang}
\affiliation{State Key Laboratory of Low-Dimensional Quantum Physics, Department of Physics, Tsinghua University, Beijing 100084, China}

\author{Meng Khoon Tey}
\email{mengkhoon\_tey@mail.tsinghua.edu.cn}
\affiliation{State Key Laboratory of Low-Dimensional Quantum Physics, Department of Physics, Tsinghua University, Beijing 100084, China}
\affiliation{Frontier Science Center for Quantum Information, Beijing, China}
\affiliation{Collaborative Innovation Center of Quantum Matter, Beijing, China}
\affiliation{Hefei National Laboratory, Hefei, Anhui 230088, China}

\date{\today}

\begin{abstract}
We report on the observation of a conspicuous loss in an ultracold mixture of $^{7}$Li and $^{88}$Sr atoms confined in a far-off-resonance optical dipole trap. We attribute the trap loss to the three-body inelastic Li-Sr-Sr collision and extract the corresponding three-body recombination coefficient $K_3$ at $T\sim 18.5,45,70,600\,\si{\micro K}$. The measured three-body recombination coefficient is about two to three orders of magnitude larger than the typical values convenient for realizing quantum degenerate gases. It also indicates a potentially large $s$-wave scattering length between the bosonic $^{7}$Li and $^{88}$Sr atoms, and essentially rules out the prospect of realizing $^7$Li and $^{88}$Sr mixtures of high phase space density.
\end{abstract}


\maketitle


\section{INTRODUCTION}
Ultracold mixtures of different atomic species provide unique opportunities to study exotic few-body and many-body physics such as Efimov states and strongly interacting quantum systems with mass imbalanced components~\cite{kohstall2012metastability,pires2014observation,maier2015efimov,maier2015efimov,barontini2009observation,tung2014geometric}. Ultracold molecules can facilitate precision measurement and test of fundamental physical laws~\cite{zelevinsky2008precision,hudson2011improved,carr2009cold}. They also benefit the study of ultracold chemistry~\cite{ospelkaus2010quantum,krems2008cold,Yang2022triatomic}. Among the possible platforms, dimers made up of alkali (AK) and alkaline-earth(-like) (AE) atoms have attracted much attention recently~\cite{PhysRevA.107.023114,PhysRevA.105.023306,C8CP03919D,PhysRevX.10.031037,PhysRevLett.108.043201,PhysRevResearch.4.043072,Schaefer_2020, Florian2018FRinRbSr,Simon2018CsYb}. Compared to the more widely studied bi-alkalis, the AK-AE molecules possess both electric and magnetic dipole moments in their absolute ground state due to an unpaired electron. Besides allowing for more controls, the extra degree of freedom is expected to usher in much richer quantum phases and phenomena. They may be used for the studies of lattice-spin models~\cite{micheli2006toolbox}, quantum magnetism~\cite{PhysRevLett.107.115301}, and exotic topological states of dipolar gas~\cite{PhysRevLett.109.025303,PhysRevLett.103.155302,PhysRevLett.103.205301,Yao2015AQD}, etc.

\begin{figure}[!ht]
	\centering
			\includegraphics[width=1\columnwidth]{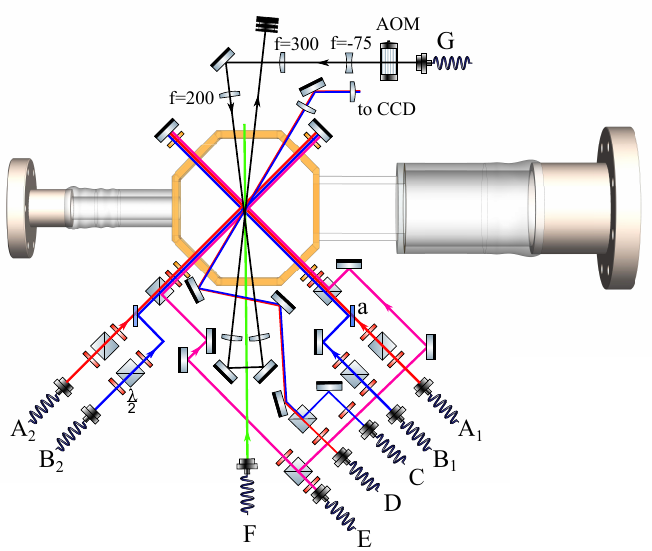}
			\caption{Optical layout around the science chamber. Zeeman slowed atomic beams enter the chamber from the left through a common dual-species Zeeman slower. A$_{1,2}$ guides the 671-nm light and 689-nm light for performing Li- and Sr red MOTs, respectively. B$_{1,2}$: 461-nm  light for Sr blue MOT, C: Sr imaging beam, D: Li imaging beam, E: grey molasses beams for Li, F: Sr repumping light (679\,nm and 707\,nm), G: 1064\,nm crossed optical dipole trap (CODT), a: dichroic beam splitter.}
			\label{fig:light-path-diagram}
\end{figure}

When preparing an ultracold mixture of AK and AE atoms, the latter can usually be laser cooled directly to much lower temperatures using a narrow $^1S_0-^3P_1$ transition~\cite{katori1999magneto,binnewies2001doppler}, and thus may be used to sympathetically cool its alkali partner~\cite{hara2011quantum,ivanov2011sympathetic,wilson2021quantum,pasquiou2013quantum}. Such potential is, nevertheless, governed ultimately by the scattering properties between the two partners. To achieve effective sympathetic cooling~\cite{modugno2001bose,mudrich2002sympathetic,anderlini2005sympathetic,pasquiou2013quantum}, a rule-of-thumb is that the ratio of elastic to inelastic collisions should be larger than 100~\cite{burke1998prospects,decarvalho1999buffer}. The interspecies interactions also affect the stability and miscibility of the gases at high phase space densities~\cite{cornish2000stable,papp2008tunable}. Therefore, understanding the scattering properties between the different components is the key to harvesting the advantages of these mixtures.

\begin{figure*}[!t]
			\includegraphics[width=2\columnwidth]{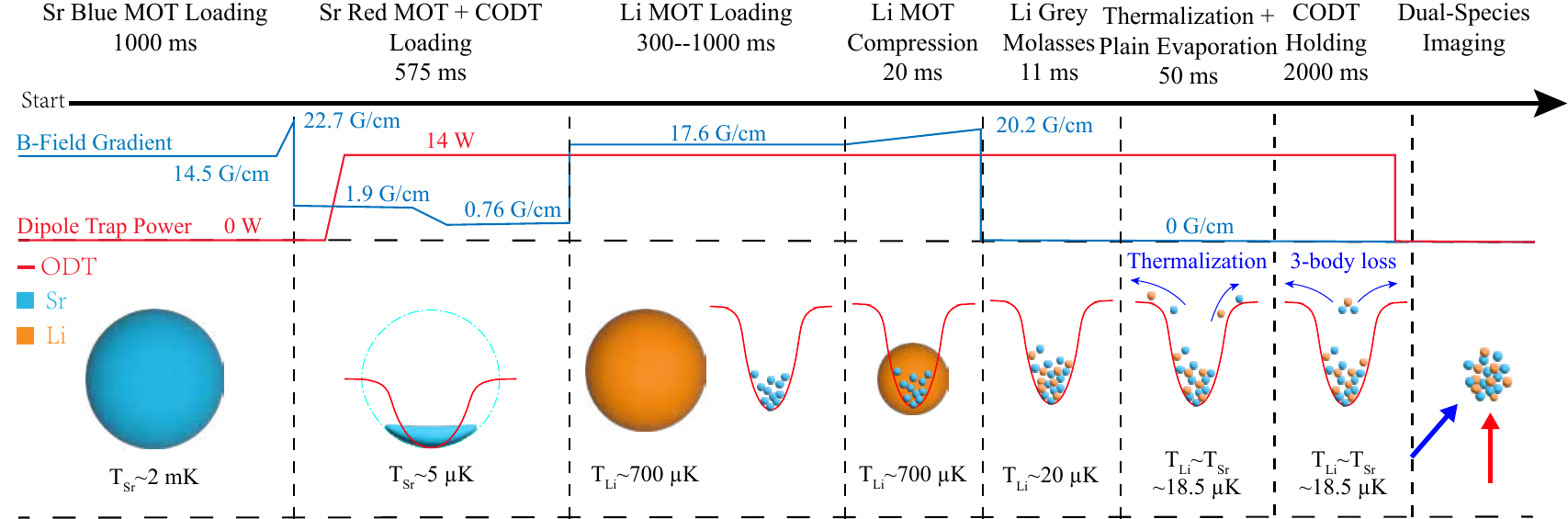}
			\caption{A typical experimental sequence. The power of dipole trap is varied to perform measurements at different temperatures. Li MOT loading time is varied to control the relative abundance between Li and Sr atoms.}
			\label{fig:time sequence}
\end{figure*}

In our previous works~\cite{xiao2019measurement,*[Erratum:]Ma2019Erratum,ye2020double}, we found that $^6$Li and various Sr isotopes have rather weak interspecies interaction. The magnitudes of their s-wave scattering lengths are of the order of 10 Bohr radii. In this work, we investigate the stability of $^7$Li and $^{88}$Sr gas mixture in an far-off-resonant optical dipole trap. We find that a $^7$Li and $^{88}$Sr mixture exhibits severe Li-Sr-Sr three-body collision loss at ultracold temperatures. By measuring the relevant three-body recombination coefficients $K_3$ at various temperatures, we conclude that the measured $K_3$ coefficients at ultracold temperatures are about two to three orders of magnitude larger than the typical values convenient for realizing quantum degenerate gases.

\section{EXPERIMENTAL PROCEDURES}\label{sec:experiment}

The optical layout around our science chamber is illustrated in Fig.~\ref{fig:light-path-diagram}. Its details have been described in our previous works~\cite{xiao2019measurement,*[Erratum:]Ma2019Erratum,ye2020double}. We measure the interspecies collision loss rate using the following experimental procedures. We begin by cooling and trapping Sr atoms in a ‘blue’ magneto-optical trap (blue MOT) using the broad 461\,nm $^{1} {S}_{0} \rightarrow {^{1} {P}_{1}}$ transition (linewidth $30.5$\,MHz). Upon completion of the blue-MOT loading, the $\sim$2\,\si{mK} atoms are further cooled in a ‘red’ MOT to 5\,\si{\micro K} using the narrow $^{1} {S}_{0} \rightarrow {^{3} {P}_{1}}$ inter-combination transition at 689\,nm (linewidth $7.5$\,kHz). During the red-MOT stage, the power of a far-off-resonant crossed optical-dipole trap (CODT) is ramped to 14\,W to store the Sr atoms. The CODT is formed by intersecting two 1064-nm light beams, both with a waist 30 $\mu$m, at 10\,\si{\degree}.

After trapping Sr atoms, Li atoms are magneto-optically trapped using the 671\,nm D$_2$ transition. To reduce loss of Sr atoms due light-assisted collision with the Li atoms, we move the Li MOT away from the Sr atoms in the CODT by changing the current of the third-stage Zeeman slower coil. Upon completion of Li atoms loading, we compress the Li cloud by increasing the magnetic field gradient and, at the same time, move the Li atoms to overlap with the CODT. Finally, we cool the Li atoms to about 20\,\si{\micro K} and transfer them into the CODT through an 11-ms-long gray-molasses (GM) cooling process. The $^7$Li atoms are pumped to the $F=1$ hyperfine ground state in the end.

The trap-depth ratio for Li and Sr is $U_{\mathrm{Li}}$/$U_{\mathrm{Sr}}$=1.1 for a 1064-nm trap. As it is much easier to obtain larger and colder sample of Sr. The Sr sympathetically cools the Li to a common temperature for the Li-Sr mixture that depends on the final trap depth of the CODT. To study the interaction properties, we hold the mixture in the CODT for a variable time and perform time-of-flight absorption imaging after switching off the CODT to determine the atom number and temperature of each species. The overall experimental sequence is detailed in Fig.~\ref{fig:time sequence}.

\section{RESULTS AND ANALYSIS}\label{sec:result}

\begin{figure}[!ht]
	\centering
			\includegraphics[width=1\columnwidth]{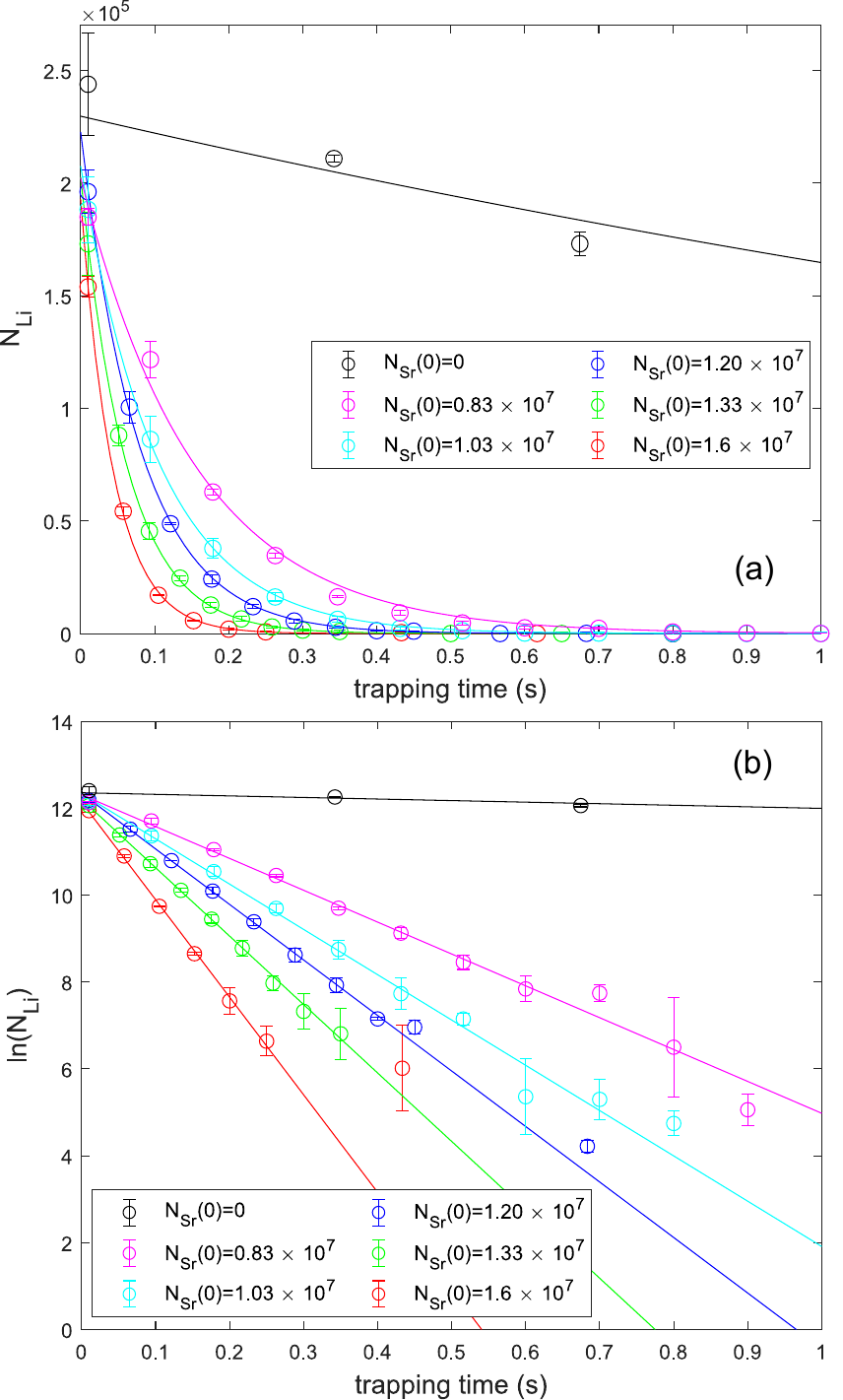}
			\caption{(a) Decay of atom numbers of $^{7}$Li in the $F=1$ hyperfine ground state in the CODT over trapping time versus a set of $^{88}$Sr atom numbers. Some error bars are smaller than the symbol size. The data are fitted with exponential functions (solid lines). (b) Same data as in (a) but using a logarithmic scale in the y axis. The solid straight lines are linear fits which highlight the single exponent nature of these decays.}
			\label{fig:Li_Decay}
\end{figure}

Figure \ref{fig:Li_Decay}(a) shows the typical loss curves of $^7$Li atoms mixed with various number of $^{88}$Sr atoms at the temperature of $T\sim600\,\si{\micro K}$. The time $t=0$ here, and in all other measurements, is chosen to be the instance when the two species become thermalized at the desired temperatures. For $T_\mathrm{Li}=T_\mathrm{Sr}=600$\,\si{\micro K}, this instance is 50\,ms after both atoms are loaded into CODT. In the absence of Sr atoms, the trapped $^{7}$Li atoms exhibit a $1/e$ lifetime of 9.1\,s, limited by collision with the background gases and plain evaporation, giving a $K_{1}\approx0.11\,$s$^{-1}$. The presence of Sr atoms results in heavy loss of $^7$Li atoms. By plotting the $^7$Li number in logarithmic scales, it is clear that all the decay curves can be very well fitted with a single-exponent $\propto \exp(-t/\tau)$, resulting in $1/e$ lifetimes $\tau$ between 40\,ms to 200\,ms.

We find that the enhanced atom loss are mainly caused by three-body recombination of Li-Sr-Sr. This conclusion is reached by considering the following loss mechanisms and observations. We first consider atom loss due to evaporative cooling. The loss rate of evaporative cooling is proportional to $\eta_{i} e^{-\eta_{i}}$~\cite{PhysRevA.53.381}, where $\eta_{i}=U_{i}/k_{B}T_{i}$ is the parameter defining the trap depth relative to thermal energy of an atom.  As we set the starting time $t=0$ such that $\eta_{\rm Li}\sim\eta_{\rm Sr}\sim10$, the single-particle atom loss caused by plain evaporation can be neglected within tens of ms. To exclude intraspecies inelastic collision loss, we measure the single-species lifetime of  Li and Sr in the CODT in the absence of another species. We find that their respective lifetimes are independent of the gas density within our parameter range, indicating a negligible intraspecies collision loss. These results are to be expected due to the small $s$-wave scattering lengths between two $F=1$ $^{7}$Li atoms ($a=39 a_0$ ~\cite{Abraham1996SingletSS,PhysRevLett.74.1315}) and between two $^{88}$Sr atoms in the ground state ($a=-2 a_0$~\cite{PhysRevLett.110.263003}).

With regard to the interspecies collision loss, the Li-Sr two-body inelastic collision is essentially forbidden due to the lack of magnetic dipole moment in the $^{1} {S}_{0}$ ground state of $^{88}$Sr. In contrast, magnetic dipole-dipole interaction is responsible for two-body dipolar spin exchange relaxation in bi-alkalis system. As for interspecies three-body recombination, two processes may occur --- the heavier recombination Li-Sr-Sr and the lighter recombination Sr-Li-Li. Typically the heavier process should dominate~\cite{PhysRevLett.103.043201,Mikkelsen_2015}. When we load much more Li than Sr ($N_{\rm Li}\sim2\times{10}^6,N_{\rm Sr}\sim5\times{10}^4$) and measure the loss rate of Sr, we do not observe enhanced Sr loss proportional to ${N_{\rm Li}}^2$. However, remarkable loss of $^7$Li is observed when we load more Sr than Li ($N_{\rm Li}\sim5\times{10}^4,N_{\rm Sr}\sim2\times{10}^6$). The above measurements thereby rule out Sr-Li-Li three-body loss as being a major source for the measurement results presented in Fig.~\ref{fig:Li_Decay} and the atom loss we observe should be attributed predominantly to the Li-Sr-Sr process.

\begin{figure}
	\centering
			\includegraphics[width=1\columnwidth]{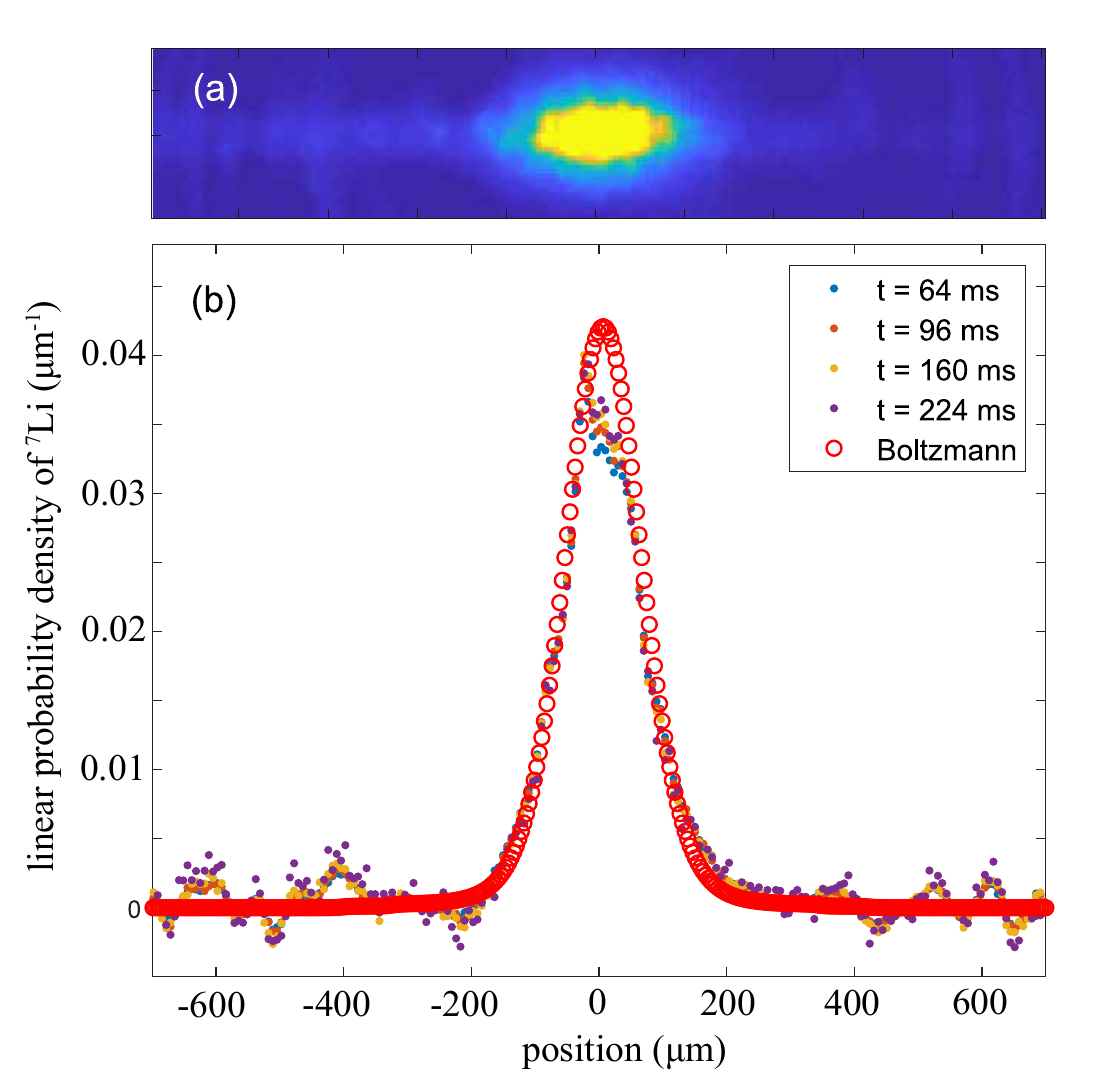}
			\caption{(a) An example of an  \textit{in situ} image of $^{7}$Li atoms in the CODT. 
(b) Probability distributions of $^{7}$Li atoms along the long axis of the CODT after various trapping times ($t=64,96,160,224$\,ms) (solid dots), obtained by dividing the measured linear density by the total atom number. Despite the atom loss over time, the probability distributions remain unchanged and their wings can be well fitted by a theoretical distribution of a thermal gas at 600\,\si{\micro K} (red line). The deviation near the cloud center is caused by the nonlinear absorption effects due to large atom densities of the order of $5\times 10^{12}$\,cm$^{-3}$.}\label{fig:Li_Boltzmann}
\end{figure}

In order to measure the three-body loss coefficient of Li-Sr-Sr, we make Sr two orders of magnitude more abundant than Li to enhance the Li-Sr-Sr process, and perform a series of experiments with different $N_{\rm Sr}$ and extract Li-Sr-Sr three-body loss coefficient by fitting the Li loss rate versus ${N_{\rm Sr}}^2$. For the mixture conditions we consider, the measured loss rate of the $^{7}$Li atoms in the Li-Sr mixtures can be described by
\begin{equation}
	\frac{dN_\mathrm{Li}}{dt} =-K_{1} N_\mathrm{Li}-K_{3} \int n_\mathrm{Li}(\vec{r})n_\mathrm{Sr}^2(\vec{r}) d^3\vec{r},
  \label{eq:Li_decay_1}
\end{equation}
where $n(\vec{r})$ represents the atom number density at position $\vec{r}$. Here, the coefficient for the single particle loss, $K_1$, is mainly determined by the amount of background gases in the vacuum chamber, whereas the coefficient for the Li-Sr-Sr three-body recombination rate, $K_{3}$, depends intrinsically on the temperature, the statistical nature and the inter-atomic potentials of the relevant atoms.

The fact that the $^7$Li decay curves are well represented by simple exponential functions is puzzling, given that the second term in Eq.~(\ref{eq:Li_decay_1}) depends on the local densities $n_\mathrm{Li}(\vec{r})$ instead of the total atom number $N_\mathrm{Li}$. It turns out that this phenomenon can be explained by the fast diffusion and re-thermalization of the Li and Sr atoms in the trap (relative to the measured $\tau$). The former is guaranteed by the high trapping frequencies and the latter can be attributed to efficient two-body scattering, a feature which usually comes hand-in-hand with fast three-body loss. As an evidence, we present in Fig.~\ref{fig:Li_Boltzmann} the measured linear probability distributions of $^{7}$Li atoms along the long axis of the CODT after certain trapping times. It is clear that the probability distributions of the $^{7}$Li atoms remain unchanged despite a large drop in the remaining atom number from $1.6\times 10^{5}$ ($t=64$\,ms) to $3.0\times 10^{4}$ ($t=224$\,ms), and can be well fitted by a Boltzmann distribution $f_\mathrm{B}(\vec{r})\propto e^{-U(\vec{r})/k_\mathrm{B}T}$, where $U(\vec{r})$ represents the trap potential. The same phenomenon is observed for the Sr.

Because of the fast thermalization behaviour, we can rewrite Eq.~(\ref{eq:Li_decay_1}) by replacing the local atom density $n_\mathrm{Li(Sr)}(\vec{r})$ by $N_\mathrm{Li(Sr)} f_\mathrm{B,Li(Sr)}(\vec{r},T)$, where $f_\mathrm{B}$ represents Boltzmann probability density distribution, giving
\begin{equation}
	\frac{dN_\mathrm{Li}}{dt} =-K_{1} N_\mathrm{Li}-K_{3} N_\mathrm{Li} N_\mathrm{Sr}^2 \mathcal{D}_\mathrm{LiSrSr},
  \label{eq:Li_decay_2}
\end{equation}
where $\mathcal{D}_\mathrm{LiSrSr}=\int_0^{\infty} f_\mathrm{B,Li}(\vec{r},T)f_\mathrm{B,Sr}^2(\vec{r},T)d^3\vec{r}$ can be calculated from the properties of the CODT as well as the temperature of the mixtures. As the measured change in the Sr atom number $N_\mathrm{Sr}$ is less than 5\% in 1 second, if we neglect the small decrease in $N_\mathrm{Sr}$, we can obtain from Eq.~(\ref{eq:Li_decay_2}) an approximate $^{7}$Li decay rate of
\begin{equation}\label{eq:tau}
1/\tau=K_{1}+K_{3} N_\mathrm{Sr}^2 \mathcal{D}_\mathrm{LiSrSr}.
\end{equation}
This formula explains why the decay of Li atoms takes a simple exponential form in Fig.~\ref{fig:Li_Decay}.

\begin{figure}
	\centering
			\includegraphics[width=1\columnwidth]{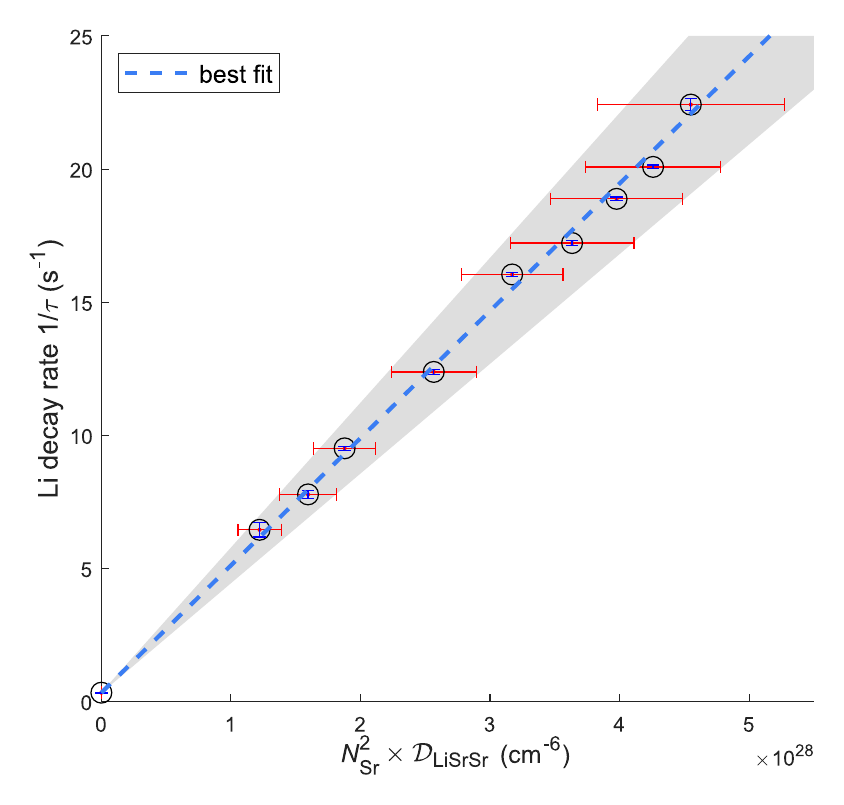}
			\caption{A least-squares linear fitting of $^{7}$Li decay rate $1/\tau$ versus $N_\mathrm{Sr}^2\times \mathcal{D}_\mathrm{LiSrSr}$. The dashed blue line is the best fit and the fitting error is represented by the gray shaded area. Some statistical error bars are smaller than the symbol size.}
			\label{fig:Linear_fitting}
\end{figure}

In Fig.~\ref{fig:Linear_fitting}, we plot the loss rate $1/\tau$ extracted from Fig.~\ref{fig:Li_Decay}(b) as a function of  $N_\mathrm{Sr}^2\mathcal{D}_\mathrm{LiSrSr}$. The resulting data fall well on a straight line in agreement with Eq.~(\ref{eq:tau}). The slope of the least-square linear fit gives a three-body recombination coefficient $K_{3}=(4.8\pm0.7)\times 10^{-28}$\,cm$^{6}$/s. This result is, nevertheless, an underestimation since we ignore the slight decay in $N_\mathrm{Sr}$. Taking into account the loss in Sr and fitting the measurement results in Fig.~\ref{fig:Li_Decay} using Eq.~(2) directly gives $K_{3}=(5.7\pm1.6)\times10^{-28}$\,cm$^{6}$/s, taking the possible systematic errors into account.

By repeating the above-mentioned procedures for $T=18.5,45,70\,\si{\micro K}$, we obtain $K_3$ at each temperature as displayed in the following table:
\begin{center}
\begin{tabular}{||c | c||}
\emph{$T/\si{\micro K}$} & \emph{$K_3/\si{cm}^{6}\cdot \si{s}^{-1}$}\\ \hline
18.5 &$ (1.3\pm0.5)\times10^{-26}$\\
46 & $(1.4\pm0.5)\times10^{-26}$\\
70 &$ (1.5\pm0.5)\times10^{-26}$\\
600 &$ (5.7\pm1.6)\times10^{-28}$\\
\end{tabular}
\end{center}

It is obvious that $K_3$ at lower temperatures ($T=18.5,45,70\,\si{\micro K}$) is much larger than that at higher temperature $T=600\,\si{\micro K}$. We attribute this to the larger scattering cross section at the lower temperatures. It is unclear why $K_3$ appears to saturate at lower temperature. We suspect this may be related to the atoms moving slower at lower temperatures and resulting in slower interspecies collisions. Unfortunately, due to the heavy atom loss during evaporative cooling, we are unable to perform measurements at even lower temperatures.


\section{CONCLUSION}\label{sec:conclusion}
In summary, we observe heavy loss in a bosonic mixture of $^{7}$Li and $^{88}$Sr atoms at $T=18.5,45,70,600\,\si{\micro K}$ in a crossed optical dipole trap and determine the three-body recombination coefficient of Li-Sr-Sr process at these temperatures. Compared to $^{87}$Rb whose three-body recombination coefficient is $(4.3\pm1.8)\times10^{-29}\si{cm}^{6}\cdot \si{s}^{-1}$ at $T=800\,\si{ nK}$~\cite{PhysRevLett.79.337}, the $K_3$ of Li-Sr-Sr is about three orders of magnitude larger at ultracold temperatures. It would be therefore challenging to realize double degenerate mixture of  $^7$Li-$^{88}$Sr. Our results indicate a potentially large $s$-wave scattering length between the bosonic $^{7}$Li and $^{88}$Sr atoms. If this scattering length is positive, it may facilitate efficient association of Li-Sr molecules by merging two optical dipole trap~\cite{PhysRevLett.130.223401}.


This work is supported by the National Natural Science Foundation of China (NSFC) (Grant No. 12234012), the National Key R\&D Program of China (Grant No. 2018YFA0306503)


\bibliographystyle{apsrev4-1}
%

\end{document}